\documentclass[smallcondensed,preprint]{svjour3}

\journalname{Hyperfine Interactions}

\usepackage[colorlinks=true, linkcolor=blue, citecolor=blue, anchorcolor=blue, urlcolor=blue]{hyperref}
\usepackage[numbers]{natbib}
\usepackage{latexsym}
\usepackage{mathptm}
\usepackage{amsmath}
\usepackage{graphicx}
\usepackage[leftcaption]{sidecap}
\usepackage{epic, eepic}
\usepackage{subfig}

\sidecaptionvpos{figure}{t}

\begin{document}

\title{Towards Antihydrogen Trapping and Spectroscopy at ALPHA}

\author{E.~Butler \and G.B.~Andresen \and M.D.~Ashkezari \and M.~Baquero-Ruiz \and W.~Bertsche \and P.D.~Bowe \and C.C.~Bray \and C.L.~Cesar \and S.~Chapman \and M.~Charlton  \and J.~Fajans \and T.~Friesen \and M.C.~Fujiwara \and D.R.~Gill \and J.S.~Hangst \and W.N.~Hardy \and R.S.~Hayano \and M.E.~Hayden \and A.J.~Humphries \and R.~Hydomako \and S.~Jonsell \and L.~Kurchaninov \and R.~Lambo \and N.~Madsen \and S.~Menary \and P.~Nolan \and K.~Olchanski \and A.~Olin \and A.~Povilus \and P.~Pusa \and F.~Robicheaux \and E.~Sarid \and D.M.~Silveira \and C.~So \and J.W.~Storey \and R.I.~Thompson \and D.P.~van~der~Werf \and D.~Wilding \and J.S.~Wurtele \and Y.~Yamazaki \\ ALPHA~Collaboration}
\authorrunning{E. Butler et al. (ALPHA)}

\institute{
		E.~Butler (\email{eoin.butler@cern.ch}) \and W.~Bertsche \and M.~Charlton \and A.J.~Humphries \and N.~Madsen \and D.P.~van~der~Werf \and D.~Wilding
		\at Department of Physics, Swansea University, Swansea SA2 8PP, United Kingdom
	\and
		G.B.~Andresen \and P.D.~Bowe \and J.S.~Hangst
		\at Department of Physics and Astronomy, Aarhus University, DK-8000 Aarhus C, Denmark
	\and
		M.D.~Ashkezari \and M.E.~Hayden
		\at Department of Physics, Simon Fraser University, Burnaby BC, V5A 1S6, Canada
	\and
		M.~Baquero-Ruiz \and C.C.~Bray \and S.~Chapman \and J.~Fajans \and A.~Povilus \and C.~So \and J.S.~Wurtele
		\at Department of Physics, University of California, Berkeley, CA 94720-7300, USA
	\and
		C.L.~Cesar \and R.~Lambo
		\at Instituto de F\'{i}sica, Universidade Federal do Rio de Janeiro, Rio de Janeiro 21941-972, Brazil
	\and
		T.~Friesen \and M.C.~Fujiwara \and R.~Hydomako \and R.I.~Thompson
		\at Department of Physics and Astronomy, University of Calgary, Calgary AB, T2N 1N4, Canada
	\and 
		M.C.~Fujiwara \and D.R.~Gill \and L.~Kurchaninov \and K.~Olchanski \and A.~Olin \and J.W.~Storey
		\at TRIUMF, 4004 Wesbrook Mall, Vancouver BC, V6T 2A3, Canada
		\pagebreak[4] 
	\and
		W.N.~Hardy
		\at Department of Physics and Astronomy, University of British Columbia, Vancouver BC, V6T 1Z4, Canada
	\and
		R.S.~Hayano
		\at Department of Physics, University of Tokyo, Tokyo 113-0033, Japan
	\and
		S.~Jonsell
		\at Fysikum, Stockholm University, SE-10609, Stockholm, Sweden	
	\and
		S.~Menary
		\at Department of Physics and Astronomy, York University, Toronto, ON, M3J 1P3, Canada
	\and 
		P.~Nolan \and P.~Pusa
		\at Department of Physics, University of Liverpool, Liverpool L69 7ZE, United Kingdom
	\and
		F.~Robicheaux
		\at Department of Physics, Auburn University, Auburn, AL 36849-5311, USA
	\and
		E.~Sarid
		\at Department of Physics, NRCN-Nuclear Research Center Negev, Beer Sheva, IL-84190, Israel	
	\and 
		D.M.~Silveira \and Y.~Yamazaki
		\at Atomic Physics Laboratory, RIKEN Advanced Science Institute, Wako, Saitama 351-0198, Japan\\
		Graduate School of Arts and Sciences, University of Tokyo, Tokyo 153-8902, Japan	
	}

	\date{Received: date / Accepted: date}
	
\maketitle{}
	
\begin{abstract}
Spectroscopy of antihydrogen has the potential to yield high-precision tests of the CPT theorem and shed light on the matter-antimatter imbalance in the Universe.
The ALPHA antihydrogen trap at CERN's Antiproton Decelerator aims to prepare a sample of antihydrogen atoms confined in an octupole-based Ioffe trap and to measure the frequency of several atomic transitions.
We describe our techniques to directly measure the antiproton temperature and a new technique to cool them to below 10~K.
We also show how our unique position-sensitive annihilation detector provides us with a highly sensitive method of identifying antiproton annihilations and effectively rejecting the  cosmic-ray background.
\end{abstract}

\section{Introduction}

Antihydrogen, the bound state of an antiproton and a positron, is the simplest pure antimatter atomic system.
The first cold (non-relativistic) antihydrogen atoms were synthesised by the ATHENA experiment in 2002 by combining antiprotons and positrons under cryogenic conditions in a Penning trap \cite{ATHENA_Nature}.
The neutral antihydrogen atoms formed were not confined by the electric and magnetic fields used to hold the antiprotons and positrons as non-neutral plasmas, but escaped to strike the matter of the surrounding apparatus and annihilate.
Detection of the coincident antiproton and positron annihilation signals was used to identify antihydrogen in these experiments.
However, before performing high-precision spectroscopy, it is highly desirable, perhaps even necessary, to confine the antihydrogen in an atomic trap.

\section{Atom Trap}

Atoms with a permanent magnetic dipole moment $\vec{\mu}$ can be trapped by exploiting the interaction of the dipole moment with an inhomogeneous magnetic field.
A three-dimensional maximum of magnetic field is not compatible with Maxwell's equations, but a minimum is.
Thus, only atoms with $\mu$ aligned antiparallel to the magnetic field (so-called `low-field seekers') can be trapped.

ALPHA creates a magnetic minimum using a variation of the Ioffe-Pritchard configuration \cite{Ioffe_Pritchard}, replacing the transverse quadrupole magnet with an octupole \cite{ALPHA_magnet}.
The octupole and the `mirror coils' that complete the trap are superconducting and are cooled to 4~K by immersing them in liquid helium.
The depth of the magnetic minimum produced is approximately 0.8~T, equivalent to a trap depth of $0.6~\mathrm{K}\times k_\mathrm{B}$ for ground state antihydrogen.

ALPHA's scheme to detect trapped antihydrogen is to quickly release trapped atoms from the atomic trap and detect their annihilation as they strike the apparatus.
Having the antihydrogen atoms escape over a short time minimises the background from cosmic rays that can mimic antihydrogen annihilations (see section \ref{sec:detector}), so the magnet systems have been designed to remove the stored energy in as short a time as possible.
The current has been measured to decay with a time constant of 9~ms for the octupole and 8.5~ms for the mirror coils.

The atom trap is superimposed on a Penning trap, which is used to confine the charged particles used in antihydrogen production.
The Penning trap electrodes are also cooled by a liquid helium reservoir and reach a temperature of approximately 7~K.
In the absence of external heating sources, the stored non-neutral plasmas should come into thermal equilibrium at this temperature.

Introduction of the multipolar transverse magnetic field modifies the confinement properties of the Penning trap.
In the most extreme case, this manifests as a `critical radius' \cite{CriticalRadiusTheory}, outside which particles can be lost from the trap simply because the magnetic field lines along which the particles move intersect the electrode walls.
Even if particles are not lost, the transverse field results in a higher rate of plasma diffusion \cite{GilsonFajans_Diffusion}.
As the plasma diffuses and expands, electrostatic potential energy is converted to thermal energy, resulting in a temperature higher than would be otherwise expected.

ALPHA chose to use an octupole instead of the prototypical quadrupole in its Ioffe trap to reduce the transverse fields close to the axis of the Penning trap, where the non-neutral plasmas are stored.
Though this choice can significantly ameliorate the undesirable effects, it does not eliminate them entirely.
Other sources of heating, notably the coupling of the particles to electronic noise \cite{NoiseTemperature}, will also increase the temperature.
This highlights the importance of direct, absolute measurements of the particle temperature to accurately determine the experimental conditions.

\section{Cooling and temperature measurements of antiprotons}

The temperature of a plasma can be determined by measuring the distribution of particles in the tail of a Boltzmann distribution - a technique common-place in non-neutral plasma physics \cite{TemperatureMeasurement}.
This measurement has the advantage of yielding the absolute temperature of the particles without recourse to supporting measurements (for example, of the density distribution), unlike measurements of the frequencies of the normal plasma modes \cite{modes}, which can only give a relative temperature change.
The plasmas typical in ALPHA have densities in the range $10^6$ to $10^8~\mathrm{cm^{-3}}$, with collision rates high enough to ensure that the plasma comes to equilibrium in a few seconds.
In equilibrium, the energy of the particles conforms to a Boltzmann distribution.

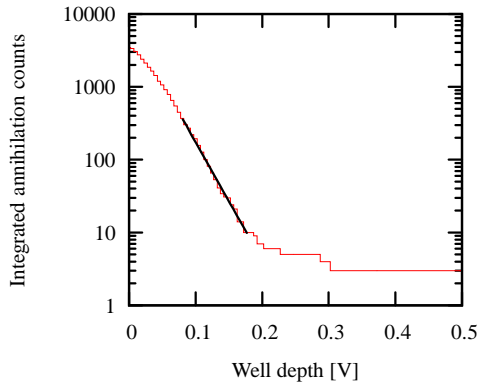
\begin{SCfigure}[1.0][h]
\centering
\setlength{\unitlength}{0.120450pt}
\begin{picture}(1650,1259)(0,0)
\footnotesize
\color{black}
\thicklines \path(513,265)(554,265)
\thicklines \path(1547,265)(1506,265)
\put(472,265){\makebox(0,0)[r]{ 1}}
\color{black}
\thicklines \path(513,334)(533,334)
\thicklines \path(1547,334)(1527,334)
\thicklines \path(513,374)(533,374)
\thicklines \path(1547,374)(1527,374)
\thicklines \path(513,402)(533,402)
\thicklines \path(1547,402)(1527,402)
\thicklines \path(513,425)(533,425)
\thicklines \path(1547,425)(1527,425)
\thicklines \path(513,443)(533,443)
\thicklines \path(1547,443)(1527,443)
\thicklines \path(513,458)(533,458)
\thicklines \path(1547,458)(1527,458)
\thicklines \path(513,471)(533,471)
\thicklines \path(1547,471)(1527,471)
\thicklines \path(513,483)(533,483)
\thicklines \path(1547,483)(1527,483)
\thicklines \path(513,493)(554,493)
\thicklines \path(1547,493)(1506,493)
\put(472,493){\makebox(0,0)[r]{ 10}}
\color{black}
\thicklines \path(513,562)(533,562)
\thicklines \path(1547,562)(1527,562)
\thicklines \path(513,602)(533,602)
\thicklines \path(1547,602)(1527,602)
\thicklines \path(513,631)(533,631)
\thicklines \path(1547,631)(1527,631)
\thicklines \path(513,653)(533,653)
\thicklines \path(1547,653)(1527,653)
\thicklines \path(513,671)(533,671)
\thicklines \path(1547,671)(1527,671)
\thicklines \path(513,686)(533,686)
\thicklines \path(1547,686)(1527,686)
\thicklines \path(513,699)(533,699)
\thicklines \path(1547,699)(1527,699)
\thicklines \path(513,711)(533,711)
\thicklines \path(1547,711)(1527,711)
\thicklines \path(513,722)(554,722)
\thicklines \path(1547,722)(1506,722)
\put(472,722){\makebox(0,0)[r]{ 100}}
\color{black}
\thicklines \path(513,790)(533,790)
\thicklines \path(1547,790)(1527,790)
\thicklines \path(513,830)(533,830)
\thicklines \path(1547,830)(1527,830)
\thicklines \path(513,859)(533,859)
\thicklines \path(1547,859)(1527,859)
\thicklines \path(513,881)(533,881)
\thicklines \path(1547,881)(1527,881)
\thicklines \path(513,899)(533,899)
\thicklines \path(1547,899)(1527,899)
\thicklines \path(513,914)(533,914)
\thicklines \path(1547,914)(1527,914)
\thicklines \path(513,928)(533,928)
\thicklines \path(1547,928)(1527,928)
\thicklines \path(513,939)(533,939)
\thicklines \path(1547,939)(1527,939)
\thicklines \path(513,950)(554,950)
\thicklines \path(1547,950)(1506,950)
\put(472,950){\makebox(0,0)[r]{ 1000}}
\color{black}
\thicklines \path(513,1018)(533,1018)
\thicklines \path(1547,1018)(1527,1018)
\thicklines \path(513,1059)(533,1059)
\thicklines \path(1547,1059)(1527,1059)
\thicklines \path(513,1087)(533,1087)
\thicklines \path(1547,1087)(1527,1087)
\thicklines \path(513,1109)(533,1109)
\thicklines \path(1547,1109)(1527,1109)
\thicklines \path(513,1127)(533,1127)
\thicklines \path(1547,1127)(1527,1127)
\thicklines \path(513,1143)(533,1143)
\thicklines \path(1547,1143)(1527,1143)
\thicklines \path(513,1156)(533,1156)
\thicklines \path(1547,1156)(1527,1156)
\thicklines \path(513,1168)(533,1168)
\thicklines \path(1547,1168)(1527,1168)
\thicklines \path(513,1178)(554,1178)
\thicklines \path(1547,1178)(1506,1178)
\put(472,1178){\makebox(0,0)[r]{ 10000}}
\color{black}
\thicklines \path(513,265)(513,306)
\thicklines \path(513,1178)(513,1137)
\put(513,182){\makebox(0,0){ 0}}
\color{black}
\thicklines \path(720,265)(720,306)
\thicklines \path(720,1178)(720,1137)
\put(720,182){\makebox(0,0){ 0.1}}
\color{black}
\thicklines \path(927,265)(927,306)
\thicklines \path(927,1178)(927,1137)
\put(927,182){\makebox(0,0){ 0.2}}
\color{black}
\thicklines \path(1133,265)(1133,306)
\thicklines \path(1133,1178)(1133,1137)
\put(1133,182){\makebox(0,0){ 0.3}}
\color{black}
\thicklines \path(1340,265)(1340,306)
\thicklines \path(1340,1178)(1340,1137)
\put(1340,182){\makebox(0,0){ 0.4}}
\color{black}
\thicklines \path(1547,265)(1547,306)
\thicklines \path(1547,1178)(1547,1137)
\put(1547,182){\makebox(0,0){ 0.5}}
\color{black}
\color{black}
\thicklines \path(513,1178)(513,265)(1547,265)(1547,1178)(513,1178)
\color{black}
\put(143,721){\makebox(0,0)[l]{\rotatebox[origin=c]{90}{Integrated annihilation counts}}}
\color{black}
\color{black}
\put(1030,58){\makebox(0,0){Well depth [V]}}
\color{black}
\color{black}
\color{red}
\thinlines \path(513,1073)(518,1073)(518,1070)(528,1070)(528,1060)(539,1060)(539,1050)(549,1050)(549,1036)(559,1036)(559,1024)(570,1024)(570,1011)(580,1011)(580,999)(590,999)(590,985)(601,985)(601,967)(611,967)(611,956)(621,956)(621,941)(632,941)(632,926)(642,926)(642,907)(652,907)(652,890)(663,890)(663,870)(673,870)(673,851)(683,851)(683,831)(694,831)(694,820)(704,820)(704,800)(714,800)(714,787)(725,787)(725,766)(735,766)(735,745)(745,745)(745,722)(756,722)(756,701)(766,701)(766,680)
\thinlines \path(766,680)(776,680)(776,659)(787,659)(787,633)(797,633)(797,615)(807,615)(807,605)(818,605)(818,602)(828,602)(828,580)(838,580)(838,567)(849,567)(849,527)(859,527)(859,527)(869,527)(869,493)(880,493)(880,493)(890,493)(890,493)(900,493)(900,483)(911,483)(911,458)(921,458)(921,458)(931,458)(931,443)(942,443)(942,443)(952,443)(952,443)(962,443)(962,443)(973,443)(973,443)(983,443)(983,425)(993,425)(993,425)(1004,425)(1004,425)(1014,425)(1014,425)(1024,425)(1024,425)
\thinlines \path(1024,425)(1035,425)(1035,425)(1045,425)(1045,425)(1055,425)(1055,425)(1066,425)(1066,425)(1076,425)(1076,425)(1086,425)(1086,425)(1097,425)(1097,425)(1107,425)(1107,402)(1117,402)(1117,402)(1128,402)(1128,402)(1138,402)(1138,374)(1148,374)(1148,374)(1159,374)(1159,374)(1169,374)(1169,374)(1179,374)(1179,374)(1190,374)(1190,374)(1200,374)(1200,374)(1210,374)(1210,374)(1221,374)(1221,374)(1231,374)(1231,374)(1241,374)(1241,374)(1252,374)(1252,374)(1262,374)(1262,374)(1272,374)(1272,374)(1283,374)(1283,374)
\thinlines \path(1283,374)(1293,374)(1293,374)(1303,374)(1303,374)(1314,374)(1314,374)(1324,374)(1324,374)(1334,374)(1334,374)(1345,374)(1345,374)(1355,374)(1355,374)(1365,374)(1365,374)(1376,374)(1376,374)(1386,374)(1386,374)(1396,374)(1396,374)(1407,374)(1407,374)(1417,374)(1417,374)(1427,374)(1427,374)(1438,374)(1438,374)(1448,374)(1448,374)(1458,374)(1458,374)(1469,374)(1469,374)(1479,374)(1479,374)(1489,374)(1489,374)(1500,374)(1500,374)(1510,374)(1510,374)(1520,374)(1520,374)(1531,374)(1531,374)(1541,374)(1541,374)
\thinlines \path(1541,374)(1547,374)(1547,374)
\color{blue}
\color{black}
\thicklines \path(680,848)(680,848)(691,829)(701,810)(711,791)(722,773)(732,754)(743,735)(753,717)(764,698)(774,679)(785,660)(795,642)(805,623)(816,604)(826,585)(837,567)(847,548)(858,529)(868,511)(879,492)
\color{black}
\thicklines \path(513,1178)(513,265)(1547,265)(1547,1178)(513,1178)
\color{black}
\end{picture}
\caption{An example temperature measurement of approximately 45,000 antiprotons, after separation from the cooling electrons and with the inhomogeneous trapping fields energised. The straight line shows an exponential fit to determine the temperature, which in this case, is $\left(310~\pm~20\right)~\mathrm{K}$}
\label{fig:temperature}
\end{SCfigure}

To measure the temperature, the particles are released from a confining well by slowly (compared to the axial oscillation frequency) reducing the voltage on one side of the well.
As the well depth is reduced, particles escape according to their energy; the first (highest-energy) particles to be released will be drawn from the tail of a Boltzmann distribution.
As the dump progresses, the loss of particles causes redistribution of energy and, at later times, the measured distribution deviates from the expected Boltzmann distribution.
The escaping particles can be detected using a micro-channel plate as a charge amplifier, or for antimatter particles, by detecting their annihilation.
The temperature is determined by fitting an exponential curve to the number of particles released as a function of energy, such as in the example measurement shown in Fig. \ref{fig:temperature}.

The actual process of manipulating the trap potentials can change the temperature of the particles as the measurement takes place.
Particle-in-cell (PIC) simulations of the measurement process have predicted that the temperature obtained from the fit is around 15\% higher than the initial temperature for a typical antiproton cloud.
For the denser electron and positron plasmas, the measured temperature can be as much as factor of two higher than the initial temperature.
We can apply the corrections determined from these simulations to the measured temperature to find the true temperature.
This temperature diagnostic has been applied to all three particle species used in ALPHA - antiprotons, positrons and electrons.
The lowest temperatures measured for electron or positron plasmas at experimentally relevant densities $\left(10^6~\mathrm{cm^{-3}} \text{or more}\right)$ is of the order of 40~K.

Electrons are used to collisionally cool the antiprotons, which, due to their larger mass, do not effectively self-cool via synchrotron radiation.
Before mixing the antiprotons with positrons to produce antihydrogen, the electrons must be removed.
If the electrons were allowed to remain, they could potentially deplete the positron plasma by forming positronium, destroy antihydrogen atoms through charge exchange, or destabilise the positron plasma by partially neutralising it.

Electron removal is accomplished through the application of electric field pulses.
These pulses remove the confining potential on one side of the well holding the antiproton/electron two-component plasma, typically for 100-300~ns.
The electrons, moving faster than the antiprotons, escape the well.
The well is restored before the antiprotons can escape, so they remain trapped.
However, the process does not avoid disturbing the antiprotons.
The electron removal process has been the focus of a significant portion of experimental effort at ALPHA, and the coldest antiproton temperatures obtained have been around 200-300~K.

\section{Evaporative Cooling}

Antiprotons at a few hundred Kelvin will have a very small probability of forming low-energy, trappable, antihydrogen atoms.
To further cool the antiprotons, ALPHA has implemented a technique of forced evaporative cooling.
Evaporative cooling is a common-place technique in neutral particle trapping, and has been instrumental in the production of Bose-Einstein condensates \cite{EVC_in_atoms}.
However, evaporative cooling has found limited application to charged particles.

Before evaporative cooling, a cloud of antiprotons, containing 45,000 particles, with a radius of 0.6~mm, density $7.6\times10^6~\mathrm{cm^{-3}}$, and initial temperature of $\left(1040~\pm~45\right)~\mathrm{K}$ was prepared in a 1.5~V deep potential well.
The collision rate between antiprotons was of order 200~$\mathrm{s}^{-1}$, high enough to ensure that the temperatures in the parallel and perpendicular degrees of freedom had equilibrated before evaporative cooling commenced.

To perform evaporative cooling, the confining potential on one side of the well is slowly (with respect to the equilibration rate) lowered.
Particles with kinetic energy higher than the instantaneous well depth escape the trap, carrying with them energy in excess of the mean thermal energy.
The distribution then evolves towards a Boltzmann distribution with lower temperature, and the process continues.

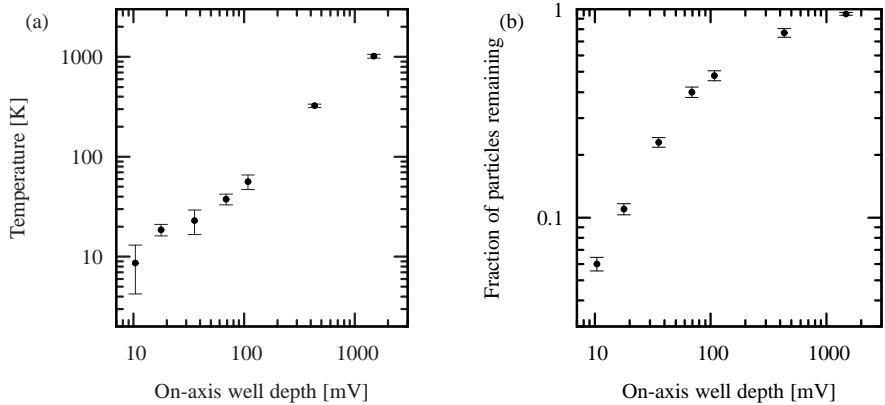
\begin{figure}[h]
\centering
\subfloat[]{
\setlength{\unitlength}{0.120450pt}
\begin{picture}(1439,1259)(0,0)
\footnotesize
\thicklines \path(431,265)(451,265)
\thicklines \path(1336,265)(1316,265)
\thicklines \path(431,320)(451,320)
\thicklines \path(1336,320)(1316,320)
\thicklines \path(431,359)(451,359)
\thicklines \path(1336,359)(1316,359)
\thicklines \path(431,390)(451,390)
\thicklines \path(1336,390)(1316,390)
\thicklines \path(431,414)(451,414)
\thicklines \path(1336,414)(1316,414)
\thicklines \path(431,435)(451,435)
\thicklines \path(1336,435)(1316,435)
\thicklines \path(431,453)(451,453)
\thicklines \path(1336,453)(1316,453)
\thicklines \path(431,469)(451,469)
\thicklines \path(1336,469)(1316,469)
\thicklines \path(431,484)(472,484)
\thicklines \path(1336,484)(1295,484)
\put(390,484){\makebox(0,0)[r]{ 10}}
\thicklines \path(431,578)(451,578)
\thicklines \path(1336,578)(1316,578)
\thicklines \path(431,633)(451,633)
\thicklines \path(1336,633)(1316,633)
\thicklines \path(431,672)(451,672)
\thicklines \path(1336,672)(1316,672)
\thicklines \path(431,703)(451,703)
\thicklines \path(1336,703)(1316,703)
\thicklines \path(431,727)(451,727)
\thicklines \path(1336,727)(1316,727)
\thicklines \path(431,748)(451,748)
\thicklines \path(1336,748)(1316,748)
\thicklines \path(431,766)(451,766)
\thicklines \path(1336,766)(1316,766)
\thicklines \path(431,782)(451,782)
\thicklines \path(1336,782)(1316,782)
\thicklines \path(431,797)(472,797)
\thicklines \path(1336,797)(1295,797)
\put(390,797){\makebox(0,0)[r]{ 100}}
\thicklines \path(431,891)(451,891)
\thicklines \path(1336,891)(1316,891)
\thicklines \path(431,946)(451,946)
\thicklines \path(1336,946)(1316,946)
\thicklines \path(431,985)(451,985)
\thicklines \path(1336,985)(1316,985)
\thicklines \path(431,1015)(451,1015)
\thicklines \path(1336,1015)(1316,1015)
\thicklines \path(431,1040)(451,1040)
\thicklines \path(1336,1040)(1316,1040)
\thicklines \path(431,1061)(451,1061)
\thicklines \path(1336,1061)(1316,1061)
\thicklines \path(431,1079)(451,1079)
\thicklines \path(1336,1079)(1316,1079)
\thicklines \path(431,1095)(451,1095)
\thicklines \path(1336,1095)(1316,1095)
\thicklines \path(431,1110)(472,1110)
\thicklines \path(1336,1110)(1295,1110)
\put(390,1110){\makebox(0,0)[r]{ 1000}}
\thicklines \path(431,1204)(451,1204)
\thicklines \path(1336,1204)(1316,1204)
\thicklines \path(431,1259)(451,1259)
\thicklines \path(1336,1259)(1316,1259)
\thicklines \path(431,265)(431,285)
\thicklines \path(431,1259)(431,1239)
\thicklines \path(451,265)(451,285)
\thicklines \path(451,1259)(451,1239)
\thicklines \path(469,265)(469,285)
\thicklines \path(469,1259)(469,1239)
\thicklines \path(484,265)(484,306)
\thicklines \path(484,1259)(484,1218)
\put(484,182){\makebox(0,0){ 10}}
\thicklines \path(588,265)(588,285)
\thicklines \path(588,1259)(588,1239)
\thicklines \path(648,265)(648,285)
\thicklines \path(648,1259)(648,1239)
\thicklines \path(691,265)(691,285)
\thicklines \path(691,1259)(691,1239)
\thicklines \path(725,265)(725,285)
\thicklines \path(725,1259)(725,1239)
\thicklines \path(752,265)(752,285)
\thicklines \path(752,1259)(752,1239)
\thicklines \path(775,265)(775,285)
\thicklines \path(775,1259)(775,1239)
\thicklines \path(795,265)(795,285)
\thicklines \path(795,1259)(795,1239)
\thicklines \path(812,265)(812,285)
\thicklines \path(812,1259)(812,1239)
\thicklines \path(828,265)(828,306)
\thicklines \path(828,1259)(828,1218)
\put(828,182){\makebox(0,0){ 100}}
\thicklines \path(932,265)(932,285)
\thicklines \path(932,1259)(932,1239)
\thicklines \path(992,265)(992,285)
\thicklines \path(992,1259)(992,1239)
\thicklines \path(1035,265)(1035,285)
\thicklines \path(1035,1259)(1035,1239)
\thicklines \path(1068,265)(1068,285)
\thicklines \path(1068,1259)(1068,1239)
\thicklines \path(1096,265)(1096,285)
\thicklines \path(1096,1259)(1096,1239)
\thicklines \path(1119,265)(1119,285)
\thicklines \path(1119,1259)(1119,1239)
\thicklines \path(1139,265)(1139,285)
\thicklines \path(1139,1259)(1139,1239)
\thicklines \path(1156,265)(1156,285)
\thicklines \path(1156,1259)(1156,1239)
\thicklines \path(1172,265)(1172,306)
\thicklines \path(1172,1259)(1172,1218)
\put(1172,182){\makebox(0,0){ 1000}}
\thicklines \path(1275,265)(1275,285)
\thicklines \path(1275,1259)(1275,1239)
\thicklines \path(1336,265)(1336,285)
\thicklines \path(1336,1259)(1336,1239)
\thicklines \path(431,1259)(431,265)(1336,265)(1336,1259)(431,1259)
\put(102,762){\makebox(0,0)[l]{\rotatebox[origin=c]{90}{Temperature [K]}}}
\put(883,58){\makebox(0,0){On-axis well depth [mV]}}
\thinlines \path(1047,951)(1047,962)
\thinlines \path(1027,951)(1067,951)
\thinlines \path(1027,962)(1067,962)
\thinlines \path(840,694)(840,740)
\thinlines \path(820,694)(860,694)
\thinlines \path(820,740)(860,740)
\thinlines \path(772,646)(772,680)
\thinlines \path(752,646)(792,646)
\thinlines \path(752,680)(792,680)
\thinlines \path(674,553)(674,630)
\thinlines \path(654,553)(694,553)
\thinlines \path(654,630)(694,630)
\thinlines \path(570,549)(570,585)
\thinlines \path(550,549)(590,549)
\thinlines \path(550,585)(590,585)
\thinlines \path(490,367)(490,520)
\thinlines \path(470,367)(510,367)
\thinlines \path(470,520)(510,520)
\thinlines \path(1231,1106)(1231,1118)
\thinlines \path(1211,1106)(1251,1106)
\thinlines \path(1211,1118)(1251,1118)
\put(1047,957){\circle*{18}}
\put(840,719){\circle*{18}}
\put(772,664){\circle*{18}}
\put(674,597){\circle*{18}}
\put(570,568){\circle*{18}}
\put(490,464){\circle*{18}}
\put(1231,1112){\circle*{18}}
\thicklines \path(431,1259)(431,265)(1336,265)(1336,1259)(431,1259)
\end{picture}}
\subfloat[]{
\setlength{\unitlength}{0.120450pt}
\begin{picture}(1439,1259)(0,0)
\footnotesize
\thicklines \path(390,265)(410,265)
\thicklines \path(1336,265)(1316,265)
\thicklines \path(390,347)(410,347)
\thicklines \path(1336,347)(1316,347)
\thicklines \path(390,410)(410,410)
\thicklines \path(1336,410)(1316,410)
\thicklines \path(390,461)(410,461)
\thicklines \path(1336,461)(1316,461)
\thicklines \path(390,505)(410,505)
\thicklines \path(1336,505)(1316,505)
\thicklines \path(390,543)(410,543)
\thicklines \path(1336,543)(1316,543)
\thicklines \path(390,576)(410,576)
\thicklines \path(1336,576)(1316,576)
\thicklines \path(390,606)(431,606)
\thicklines \path(1336,606)(1295,606)
\put(349,606){\makebox(0,0)[r]{ 0.1}}
\thicklines \path(390,803)(410,803)
\thicklines \path(1336,803)(1316,803)
\thicklines \path(390,918)(410,918)
\thicklines \path(1336,918)(1316,918)
\thicklines \path(390,999)(410,999)
\thicklines \path(1336,999)(1316,999)
\thicklines \path(390,1063)(410,1063)
\thicklines \path(1336,1063)(1316,1063)
\thicklines \path(390,1114)(410,1114)
\thicklines \path(1336,1114)(1316,1114)
\thicklines \path(390,1158)(410,1158)
\thicklines \path(1336,1158)(1316,1158)
\thicklines \path(390,1196)(410,1196)
\thicklines \path(1336,1196)(1316,1196)
\thicklines \path(390,1229)(410,1229)
\thicklines \path(1336,1229)(1316,1229)
\thicklines \path(390,1259)(431,1259)
\thicklines \path(1336,1259)(1295,1259)
\put(349,1259){\makebox(0,0)[r]{ 1}}
\thicklines \path(390,265)(390,285)
\thicklines \path(390,1259)(390,1239)
\thicklines \path(411,265)(411,285)
\thicklines \path(411,1259)(411,1239)
\thicklines \path(429,265)(429,285)
\thicklines \path(429,1259)(429,1239)
\thicklines \path(446,265)(446,306)
\thicklines \path(446,1259)(446,1218)
\put(446,182){\makebox(0,0){ 10}}
\thicklines \path(554,265)(554,285)
\thicklines \path(554,1259)(554,1239)
\thicklines \path(617,265)(617,285)
\thicklines \path(617,1259)(617,1239)
\thicklines \path(662,265)(662,285)
\thicklines \path(662,1259)(662,1239)
\thicklines \path(697,265)(697,285)
\thicklines \path(697,1259)(697,1239)
\thicklines \path(725,265)(725,285)
\thicklines \path(725,1259)(725,1239)
\thicklines \path(749,265)(749,285)
\thicklines \path(749,1259)(749,1239)
\thicklines \path(770,265)(770,285)
\thicklines \path(770,1259)(770,1239)
\thicklines \path(789,265)(789,285)
\thicklines \path(789,1259)(789,1239)
\thicklines \path(805,265)(805,306)
\thicklines \path(805,1259)(805,1218)
\put(805,182){\makebox(0,0){ 100}}
\thicklines \path(913,265)(913,285)
\thicklines \path(913,1259)(913,1239)
\thicklines \path(977,265)(977,285)
\thicklines \path(977,1259)(977,1239)
\thicklines \path(1021,265)(1021,285)
\thicklines \path(1021,1259)(1021,1239)
\thicklines \path(1056,265)(1056,285)
\thicklines \path(1056,1259)(1056,1239)
\thicklines \path(1085,265)(1085,285)
\thicklines \path(1085,1259)(1085,1239)
\thicklines \path(1109,265)(1109,285)
\thicklines \path(1109,1259)(1109,1239)
\thicklines \path(1130,265)(1130,285)
\thicklines \path(1130,1259)(1130,1239)
\thicklines \path(1148,265)(1148,285)
\thicklines \path(1148,1259)(1148,1239)
\thicklines \path(1165,265)(1165,306)
\thicklines \path(1165,1259)(1165,1218)
\put(1165,182){\makebox(0,0){ 1000}}
\thicklines \path(1273,265)(1273,285)
\thicklines \path(1273,1259)(1273,1239)
\thicklines \path(1336,265)(1336,285)
\thicklines \path(1336,1259)(1336,1239)
\thicklines \path(390,1259)(390,265)(1336,265)(1336,1259)(390,1259)
\put(102,762){\makebox(0,0)[l]{\rotatebox[origin=c]{90}{Fraction of particles remaining}}}
\put(863,58){\makebox(0,0){On-axis well depth [mV]}}
\thinlines \path(1226,1240)(1226,1249)
\thinlines \path(1206,1240)(1246,1240)
\thinlines \path(1206,1249)(1246,1249)
\thinlines \path(817,1035)(817,1066)
\thinlines \path(797,1035)(837,1035)
\thinlines \path(797,1066)(837,1066)
\thinlines \path(536,615)(536,650)
\thinlines \path(516,615)(556,615)
\thinlines \path(516,650)(556,650)
\thinlines \path(747,983)(747,1015)
\thinlines \path(727,983)(767,983)
\thinlines \path(727,1015)(767,1015)
\thinlines \path(644,827)(644,857)
\thinlines \path(624,827)(664,827)
\thinlines \path(624,857)(664,857)
\thinlines \path(1034,1171)(1034,1198)
\thinlines \path(1014,1171)(1054,1171)
\thinlines \path(1014,1198)(1054,1198)
\thinlines \path(452,439)(452,482)
\thinlines \path(432,439)(472,439)
\thinlines \path(432,482)(472,482)
\put(1226,1244){\circle*{18}}
\put(817,1051){\circle*{18}}
\put(536,633){\circle*{18}}
\put(747,999){\circle*{18}}
\put(644,842){\circle*{18}}
\put(1034,1185){\circle*{18}}
\put(452,461){\circle*{18}}
\thicklines \path(390,1259)(390,265)(1336,265)(1336,1259)(390,1259)
\end{picture}}
\caption{The temperature (a) and the fraction of the initial number of particles (b) after evaporative cooling to a series of well depths. The minimum temperature is (9 $\pm$ 4)~K}
\label{fig:EVC_data}
\end{figure}

Starting with $45,000$ antiprotons at 1040~K, we have obtained temperatures as low as (9~$\pm$~4)~K with $\left(6\pm1\right)\%$ of the particles stored in a 10~mV deep well.
Measurements of the temperature, number of particles and transverse size of the clouds were made at a number of points between the most extreme well depths.
The temperatures and number of particles remaining at each measurement point are shown in Fig. \ref{fig:EVC_data}.

\captionsetup[subfloat]{position=top,captionskip=-10pt, justification=raggedright, singlelinecheck=false, margin=20pt}

\vspace{-0.5cm}

The evaporation process can be described using simple rate equations for the number of particles $N$ and the temperature $T$;

\vspace{-0.2cm}
\begin{subequations}
\begin{center}
\begin{tabular}{p{0.4\textwidth} p{0.4\textwidth}}
	\begin{equation}
		\frac{\mathrm{d}N}{\mathrm{d}t} = - \frac{N}{\tau_{ev}}, 
	\end{equation} 	&
	\begin{equation}
		\frac{\mathrm{d}T}{\mathrm{d}t} = - \alpha \frac{T}{\tau_{ev}} .
	\end{equation}
\end{tabular}
\end{center}
\end{subequations}
\vspace{-0.5cm}

\noindent Here, $\tau_{ev}$ is the characteristic evaporation timescale and $\alpha$ is the excess energy carried away by an evaporating particle, in multiples of $k_\mathrm{B} T$.
At a given time, the distribution of energies can be thought of as a truncated Boltzmann distribution, characterised by a temperature $T$, and the well depth $U$.
$\tau_{ev}$ is linked to the mean time between collisions, $\tau_{col}$ as \cite{EVC_Theory}
\begin{equation}
	\frac{\tau_{ev}}{\tau_{col}} = \frac{\sqrt{2}}{3} \eta e^\eta,
	\label{eqn:tau}
\end{equation}
where $\eta = U/{k_\mathrm{B}T}$ is the rescaled well depth.
We note the strong dependence of $\tau_{ev}$ on $\eta$, indicating that this is the primary factor determining the temperature in a given well. 
We find values of $\eta$ between 10 and 20 over the range of our measurements.
The value of $\alpha$ can be calculated using the treatment in reference \cite{ketterleReview}.
We have numerically modelled evaporative cooling in our experiment using these equations and have found very good agreement between our measurements and the model \cite{ALPHA_EVC}.

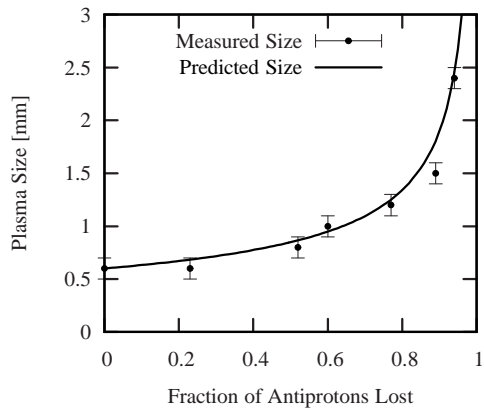
\begin{SCfigure}[1.0][h]
\setlength{\unitlength}{0.120450pt}
\begin{picture}(1650,1259)(0,0)
\footnotesize
\thicklines \path(390,265)(431,265)
\thicklines \path(1546,265)(1505,265)
\put(349,265){\makebox(0,0)[r]{ 0}}
\thicklines \path(390,431)(431,431)
\thicklines \path(1546,431)(1505,431)
\put(349,431){\makebox(0,0)[r]{ 0.5}}
\thicklines \path(390,596)(431,596)
\thicklines \path(1546,596)(1505,596)
\put(349,596){\makebox(0,0)[r]{ 1}}
\thicklines \path(390,762)(431,762)
\thicklines \path(1546,762)(1505,762)
\put(349,762){\makebox(0,0)[r]{ 1.5}}
\thicklines \path(390,928)(431,928)
\thicklines \path(1546,928)(1505,928)
\put(349,928){\makebox(0,0)[r]{ 2}}
\thicklines \path(390,1093)(431,1093)
\thicklines \path(1546,1093)(1505,1093)
\put(349,1093){\makebox(0,0)[r]{ 2.5}}
\thicklines \path(390,1259)(431,1259)
\thicklines \path(1546,1259)(1505,1259)
\put(349,1259){\makebox(0,0)[r]{ 3}}
\thicklines \path(390,265)(390,306)
\thicklines \path(390,1259)(390,1218)
\put(390,182){\makebox(0,0){ 0}}
\thicklines \path(621,265)(621,306)
\thicklines \path(621,1259)(621,1218)
\put(621,182){\makebox(0,0){ 0.2}}
\thicklines \path(852,265)(852,306)
\thicklines \path(852,1259)(852,1218)
\put(852,182){\makebox(0,0){ 0.4}}
\thicklines \path(1084,265)(1084,306)
\thicklines \path(1084,1259)(1084,1218)
\put(1084,182){\makebox(0,0){ 0.6}}
\thicklines \path(1315,265)(1315,306)
\thicklines \path(1315,1259)(1315,1218)
\put(1315,182){\makebox(0,0){ 0.8}}
\thicklines \path(1546,265)(1546,306)
\thicklines \path(1546,1259)(1546,1218)
\put(1546,182){\makebox(0,0){ 1}}
\thicklines \path(390,1259)(390,265)(1546,265)(1546,1259)(390,1259)
\put(102,762){\makebox(0,0)[l]{\rotatebox[origin=c]{90}{Plasma Size [mm]}}}
\put(968,58){\makebox(0,0){Fraction of Antiprotons Lost}}
\put(1005,1177){\makebox(0,0)[r]{Measured Size}}
\thinlines \path(1046,1177)(1251,1177)
\thinlines \path(1046,1197)(1046,1157)
\thinlines \path(1251,1197)(1251,1157)
\thinlines \path(1477,1027)(1477,1093)
\thinlines \path(1457,1027)(1497,1027)
\thinlines \path(1457,1093)(1497,1093)
\thinlines \path(1419,729)(1419,795)
\thinlines \path(1399,729)(1439,729)
\thinlines \path(1399,795)(1439,795)
\thinlines \path(1280,629)(1280,696)
\thinlines \path(1260,629)(1300,629)
\thinlines \path(1260,696)(1300,696)
\thinlines \path(1084,563)(1084,629)
\thinlines \path(1064,563)(1104,563)
\thinlines \path(1064,629)(1104,629)
\thinlines \path(991,497)(991,563)
\thinlines \path(971,497)(1011,497)
\thinlines \path(971,563)(1011,563)
\thinlines \path(656,431)(656,497)
\thinlines \path(636,431)(676,431)
\thinlines \path(636,497)(676,497)
\thinlines \path(390,431)(390,497)
\thinlines \path(370,431)(410,431)
\thinlines \path(370,497)(410,497)
\put(1477,1060){\circle*{18}}
\put(1419,762){\circle*{18}}
\put(1280,663){\circle*{18}}
\put(1084,596){\circle*{18}}
\put(991,530){\circle*{18}}
\put(656,464){\circle*{18}}
\put(390,464){\circle*{18}}
\put(1148,1177){\circle*{18}}
\put(1005,1094){\makebox(0,0)[r]{Predicted Size}}
\thicklines \path(1046,1094)(1251,1094)
\thicklines \path(390,464)(390,464)(402,465)(413,466)(425,467)(437,468)(448,469)(460,470)(472,471)(483,472)(495,474)(507,475)(518,476)(530,477)(542,478)(553,480)(565,481)(577,482)(589,483)(600,485)(612,486)(624,488)(635,489)(647,490)(659,492)(670,493)(682,495)(694,497)(705,498)(717,500)(729,501)(740,503)(752,505)(764,507)(775,508)(787,510)(799,512)(810,514)(822,516)(834,518)(845,520)(857,523)(869,525)(880,527)(892,529)(904,532)(915,534)(927,537)(939,539)(950,542)(962,545)
\thicklines \path(962,545)(974,548)(986,551)(997,554)(1009,557)(1021,560)(1032,563)(1044,567)(1056,570)(1067,574)(1079,578)(1091,582)(1102,586)(1114,590)(1126,595)(1137,599)(1149,604)(1161,609)(1172,615)(1184,620)(1196,626)(1207,632)(1219,639)(1231,646)(1242,653)(1254,661)(1266,669)(1277,677)(1289,687)(1301,697)(1312,707)(1324,719)(1336,731)(1347,745)(1359,760)(1371,776)(1383,794)(1394,814)(1406,836)(1418,861)(1429,891)(1441,924)(1453,964)(1464,1013)(1476,1073)(1488,1150)(1499,1254)(1500,1259)
\thicklines \path(390,1259)(390,265)(1546,265)(1546,1259)(390,1259)
\end{picture}
	\caption{The measured size of the antiproton cloud using a MCP/phosphor/CCD device as a function of the number of particles lost. This is compared to the size predicted from eqn \ref{eq:expansion}}
	\label{fig:radius}
\end{SCfigure}

Measurements of the transverse density profile were made by ejecting the particles onto an MCP/phosphor/CCD imaging device \cite{ALPHA_MCP}.
It was seen that, as evaporation progressed, the cloud radius increased dramatically - see Fig. \ref{fig:radius}.
We interpret this effect to be due to escape of the evaporating particles principally from the radial centre of the cloud, and the conservation of the total canonical angular momentum during the subsequent redistribution process.
Inside the cloud, the space charge reduces the depth of the confining well.
This effect is accentuated closer to the trap axis, with the result that the well depth close to the axis can be significantly lower than further away.
The evaporation rate is exponentially suppressed at higher well depths (eqn. \ref{eqn:tau}), so evaporation is confined to a small region close to the axis, causing the on-axis density to become depleted.
This is a non-equilibrium configuration, and the particles will redistribute to replace the lost density.
In doing so, some particles will move inwards, and to conserve the canonical angular momentum, some particles must also move to higher radii \cite{confinementTheorem}.
Assuming that all loss occurs at $r=0$, the mean squared radius of the particles, $\left< r^2 \right>$, will obey the relationship
\vspace{-0.5cm}
\begin{equation}
	\label{eq:expansion}
	N_0\left< r_0^2\right> = N \left< r^2 \right>,
	\vspace{-0.5cm}
\end{equation}
where N is the number of particles, and the zero subscript indicates the initial conditions.

As seen in Fig. \ref{fig:radius}, this model agrees very well with the measurements.
This radial expansion can be problematic when attempting to prepare low kinetic energy antiprotons to produce trappable antihydrogen atoms, as the energy associated with the magnetron motion grows with the distance from the axis, and the electrostatic potential energy released as the radius expands can reheat the particles.
The effect can be countered somewhat by taking a longer time to cool the particles, resulting in a higher efficiency and, thus, a smaller expansion, but we find that the efficiency depends very weakly on the cooling time.

Colder antiprotons are of great utility in the effort to produce cold antihydrogen atoms.
Antihydrogen production techniques can be broadly categorised as `static' - in which a cloud of antiprotons is held stationary and positrons, perhaps in the form of positronium atoms are introduced \cite{positronium}, or `dynamic' - where antiprotons are passed through a positron plasma \cite{Nested}.
In the first case, the advantages of cold antiprotons are obvious, as the lower kinetic energy translates directly into lower-energy antihydrogen atoms.
In the second case, the colder temperature allows the manipulations used to `inject' the antiprotons into the positrons to produce much more precisely defined antiproton energies.
Indirectly, this will also permit these schemes to produce more trappable antihydrogen.

\section{Annihilation vertex detector}\label{sec:detector}

Among the most powerful diagnostic tools available to experiments working with antimatter are detectors capable of detecting matter-antimatter annihilations.
Antiproton annihilations produce an average of three charged pions, which can be detected by scintillating material placed around the trap.
The passage of a pion through the scintillator produces photons, which trigger a cascade in a photo-multiplier tube to produce a voltage pulse.
Individual voltage pulses can be counted to determine the number of annihilations.

A further technique uses a position-sensitive detector to reconstruct the trajectories of the pions and find the point where the antiproton annihilated (usually called the `vertex').
The ALPHA annihilation vertex detector comprises sixty double-sided silicon wafers, arranged in three layers in a cylindrical fashion around the antihydrogen production and trapping region.
Each wafer is divided into 256 strips, oriented in orthogonal directions on the p- and n- sides.
Charged particles passing through the silicon result in charge deposits, and the intersection of perpendicular strips with charge above a defined threshold marks the location a particle passed through the silicon.

\begin{SCfigure}[1.0][h]
\includegraphics[width=0.6\textwidth]{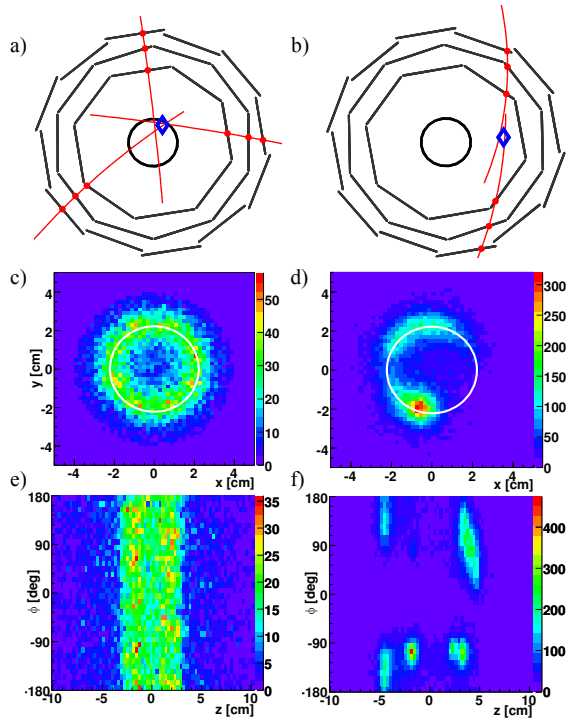}
\caption{(a) an example reconstruction of an antihydrogen annihilation and (b) a cosmic ray event. The diamond indicates the position of the vertex identified by the reconstruction algorithm, the polygonal structure shows the locations of the silicon wafers, the dots are the positions of the detected hits, and the inner circle shows the radius of the Penning trap electrodes. Also shown are annihilation density distributions associated with antihydrogen production (c, e) and deliberately induced antiproton loss (d, f). (c) and (d) are projected along the cylindrical axis, with the inner radius of the electrodes marked with a white circle, while (e) and (f) show the azimuthal angle $\phi$ against the axial position $z$}
\label{fig:vertex}
\end{SCfigure}

Each module is controlled by a circuit that produces a digital signal when a charge is detected on the silicon.
If a coincidence of modules is satisfied in a 400~ns time window, the charge profile is `read-out' and digitised for further analysis.
Each readout and associated trigger and timing information comprises an `event'.
The pion trajectories are reconstructed by fitting helices to sets of three hits, one from each layer of the detector.
The point that minimises the distance to the helices is then identified as the annihilation vertex.
An example of an annihilation event is shown in Fig. \ref{fig:vertex}(a).

\vspace{-0.5cm}
\vspace{-0.5cm}

Examination of the spatial distributions of annihilations can yield much insight into the physical processes at work.
ATHENA established that antihydrogen production resulted in a characteristic `ring' structure - an azimuthally smooth distribution concentrated at the radius of the trap electrodes \cite{ATHENA_imaging}, shown in \ref{fig:vertex}(c) and (e).
In contrast, the loss of bare antiprotons occurred in spatially well-defined locations, called `hot-spots', examples of which are shown in \ref{fig:vertex}(d) and (f).
This was interpreted to be due to microscopic imperfections in the trap elements.
These produce electric fields that break the symmetry of the trap and give rise to preferred locations for charged particle loss.
When antihydrogen is produced in a multipole field, antiprotons generated by ionisation of weakly-bound antihydrogen also contribute small asymmetries \cite{ALPHA_HbarOct}.
These features are present in Fig. \ref{fig:vertex}(c) and (e).

The vertex detector is also sensitive to charged particles in cosmic rays.
When passing through the detector, they are typically identified as a pair of almost co-linear tracks (Fig. \ref{fig:vertex}(b)), and can be misidentified as an annihilation.
Cosmic-ray events when searching for the release of trapped antihydrogen thus present a background.

\begin{SCfigure}[1.0][h]
\includegraphics[width=0.6\textwidth]{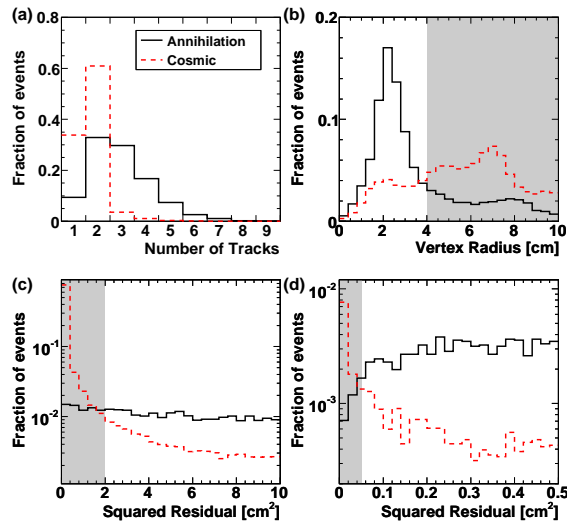}
\caption{Comparison of the distributions of event parameters for antiproton annihilations (solid line) and cosmic rays (dashed line). Shown are (a) the number of identified charged particle tracks, (b) the radial coordinate of the vertex, and the squared residual from a linear fit to the identified positions for the events with (c) two tracks and (d) more than two tracks. The shaded regions indicate the range of parameters that are rejected to minimise the p-value as discussed in the text}
\label{fig:distributions}
\end{SCfigure}

To develop a method to reject cosmic ray events, while retaining annihilations, we compared samples of the events using three parameters, shown in Fig. \ref{fig:distributions}.
Cosmic rays have predominantly two tracks, while antiproton annihilations typically have more. 95\% of cosmic events have two or fewer identified tracks, while 58\% of antiproton annihilations have at least three.
A significant number of antiproton annihilations can have only two tracks, so it is not desirable to reject all these events as background.

\vspace{-0.2cm}
\vspace{-0.4cm}

We determine if the tracks form a straight line by fitting a line to the hits from each pair of tracks, and calculating the squared residual value.
As seen in Fig. \ref{fig:distributions}(c) and (d), cosmic events have much lower squared residual values than annihilations.
This is to be expected, since particles from cosmic rays have high momentum and pass through the apparatus and the magnetic field essentially undeflected, while the particles produced in an annihilation will, in general, move in all directions.
In addition, annihilations occur on the inner wall of the Penning trap, at a radius of $\sim$2.2~cm, and as shown in Fig. \ref{fig:distributions}(b), reconstructed annihilation vertices are concentrated here, whereas cosmic rays pass through at a random radius.

By varying the ranges of parameters for which events are accepted, we could optimise the annihilation detection strategy.
The point where the `p-value' -- the probability that an observed signal is due to statistical fluctuations in the background \cite{PDG} --  was minimised requiring the vertex to lie within 4~cm of the trap axis, and the squared residual value to be at least 2~$\mathrm{cm}^2$ or 0.05~$\mathrm{cm}^2$ for events with two tracks and more than two tracks, respectively.

These thresholds reject more than 99\% of the cosmic background, reducing the absolute rate of background events to 22~mHz, while still retaining the ability of identify $\sim 40\%$ of antiproton annihilations.
While this method effectively removes cosmic rays as a source of concern, other background processes, including mirror-trapped antiprotons must also be considered when searching for trapped antihydrogen.
Our cosmic-ray rejection method has been applied to data taken from the 2009 ALPHA antihydrogen trapping run, and a full discussion of the results obtained will be made in a forthcoming publication.

\section{Conclusions and outlook}
In this paper we have described two of the most recent techniques developed by the ALPHA collaboration in our search for trapped antihydrogen.
Evaporative cooling of antiprotons has the potential to greatly increase the number of low-energy, trappable atoms produced in our experiment.
The use of our unique annihilation vertex imaging detector to discriminate with high power between annihilations and cosmic rays will be a vital tool to identify the first trapped antihydrogen atoms.
We have integrated both of these techniques into our experiment and are hopeful of soon being able to report detection of trapped antihydrogen.

\begin{acknowledgements}
This work was supported by CNPq, FINEP/RENAFAE (Brazil), ISF (Israel), MEXT (Japan), FNU (Denmark), VR (Sweden), NSERC, NRC/TRIUMF, AIF (Canada), DOE, NSF (USA), EPSRC and the Leverhulme Trust (UK).
We are also grateful to the AD team for the delivery of a high-quality antiproton beam, and to CERN for its technical support.
\end{acknowledgements}

\subsection*{Note added in proof}: Since the preparation of this article, trapping of antihydrogen atoms has been
achieved by the ALPHA collaboration \cite{ALPHA_Nature}

	\bibliographystyle{aipnum4-1}
	\bibliography{theBibliography}

\end{document}